\documentclass[aps,prd,groupedaddress,showpacs,nofootinbib,twocolumn]{revtex4}
\usepackage{graphicx, amssymb}

\newcommand{\be}{\begin{equation}}
\newcommand{\ee}{\end{equation}}
\newcommand{\bea}{\begin{eqnarray}}
\newcommand{\eea}{\end{eqnarray}}
\newcommand{\beaa}{\begin{eqnarray*}}
\newcommand{\eeaa}{\end{eqnarray*}}

\newcommand{\nn}{\nonumber \\}

\newcommand{\e}{{\rm e}}

\begin{document}
\title{Cosmological viability of  $f(R)$\,-\,gravity as an ideal fluid and
its compatibility with a matter dominated phase}
\author{S. Capozziello$^1$\footnote{Electronic address: capozziello@na.infn.it}, S. Nojiri$^2$\footnote{Electronic address: nojiri@phys.nagoya-u.ac.jp,\\ snojiri@yukawa.kyoto-u.ac.jp},
S.D. Odintsov$^3$\footnote{Electronic address:
odintsov@ieec.uab.es also at Lab. Fundamental Studies, Tomsk State
Pedagogical University, Tomsk}, A. Troisi$^1$\footnote{Electronic
address: antro@na.infn.it}}

\affiliation{$^1$Dipartimento di Scienze Fisiche, Univ. di Napoli "Federico II", and INFN, Sez. di Napoli,
Compl. Univ. di Monte S. Angelo, Ed. N, via Cinthia, 80126 - Napoli, Italy \\
$^2$ Department of Physics,
Nagoya University, Nagoya 464-8602, Japan \\
$^3$ Instituci\`{o} Catalana de Recerca i Estudis Avan\c{c}ats
(ICREA) and Institut de Ciencies de l'Espai (IEEC-CSIC),
Campus UAB, Facultat Ciencies, Torre C5-Par-2a pl, E-08193 Bellaterra
(Barcelona), Spain}
\date{\today}

\begin{abstract}
We show that $f(R)$-gravity can, in general, give rise to
cosmological viable models  compatible with a matter-dominated
epoch evolving into a late accelerated phase. We  discuss the
various representations of $f(R)$-gravity as an ideal fluid or a
scalar-tensor gravity theory, taking into account conformal
transformations. We point out that mathematical equivalence does
not correspond, in several cases,  to the physical equivalence of
Jordan frame and Einstein frame. Finally, we show that wide
classes of $f(R)$ gravity models, including matter and accelerated
phases, can be phenomenologically reconstructed by means of
observational data. In principle, any popular quintessence models
could be "reframed" as an $f(R)$-gravity model.

\end{abstract}


%\pacs{04.50.+h, 04.25.Nx, 98.80.-k}

\maketitle

\noindent 1. Up to ten years ago, the  paradigm of cosmology has
been the Cosmological Standard Model inspired by Einstein's
General Relativity (GR), the Big Bang model and particle physics.
Recently, several observational surveys have provided high quality
data  giving rise to the so called {\it precision cosmology} by
which the nowadays picture of the universe results radically
different from the standard Einstein-Friedman model. Specifically,
the Hubble diagram of Type Ia Supernovae \cite{riess},
anisotropies in the cosmic microwave background radiation
\cite{spergel}, matter power-spectrum measured in optical surveys
of large scale structures \cite{cole}  have convincingly given the
picture of a spatially flat universe in a phase of accelerated
expansion (for a review see \cite{bagla}). The new theoretical
scenario seems consistent only if huge amounts of dark components,
usually treated as fluids, are introduced into the game
\cite{padh,sami}. In particular, in order  to drive the observed
cosmic acceleration and to fill the gap between the matter-energy
content and the critical density of a spatially flat universe, a
negative pressure fluid, not clustered at small scales, is needed.
The dramatic result is that this mysterious and unexpected
component,  referred to as {\it Dark Energy}, adds to the already
supposed presence of the {\it Dark Matter} necessary to fit the
astrophysical data at smaller clustered scales. In the simplest
scenario (Concordance $\Lambda$\,-\,CDM model) \cite{concordance},
dark energy is interpreted as the cosmological constant,
contributing for ~ 70\% to the whole energy budget, while the
remaining 30\% is constituted for ~ 4\% of baryons and for ~ 25\%
of cold dark matter. Such a kind of dark matter is typically
addressed to exotic particles such as, e.g., WIMPs or axions,
nevertheless, up to now, no definitive proof of their existence
has been achieved \cite{krauss}. This model, although satisfactory
from an observational point of view, is anyway theoretically
disfavored since there is a huge ($\sim$120 orders of magnitude)
difference between the predicted and observed values of $\Lambda$.
In order to overcome the cosmological constant problem, many
alternative models have been proposed. Unfortunately, none of
these attempts is fully satisfactory, since they mostly lead to
`ad hoc' cosmologies, not well  theoretically and/or
observationally founded (for a review  see\cite{woodard,NO-rev}).

 A different approach to face
this challenge can be pursued reversing the problem and
considering extended modified gravity compatible with the
observational results. This scheme possesses the relevant feature
that experimental data turn out to be naturally interpreted
without the need of additional components/fluids. Of course,
modifying the gravity from the Einstein GR implies several
theoretical issues. In particular, one has to be able to match GR
prescriptions, at least at the post-parameterized-Newtonian limit
\cite{will}, and recover the positive and well-established results
of standard cosmology. The attempt of extending GR is indeed
theoretically and physically appealing. In fact, rather than
formulating a different approach radically changing the rules on
the standard one, one can straightforwardly extend the properties
of Einstein gravity. In such a way, the theoretical foundations
remain the same, but the search for a different theory of gravity
can be essentially pursued by inferring physical features directly
from the data. Actually, among Extended Theories of Gravity,
$f(R)$\,-\,gravity \cite{NO-rev,kerner,schmidt} represents a
viable alternative to dark energy and naturally gives rise to
accelerating singularity-free solutions in early and late cosmic
epochs. \cite{starobinsky,capoz-curv,CDTT,NOPRD,ON-curv}.
Furthermore, it has to be remarked that higher-order terms in the
gravity Lagrangian are required in theories attempting to
formulate a satisfactory quantum field theory on curved
space-times \cite{birrell} and can descend from the low energy
limit of String/M-theory \cite{ON-Mth}. These theories fit quite
well the data on the SNeIa Hubble diagram \cite{noi-ijmpd} and can
also give interesting theoretical predictions with respect to the
CMBR observational results \cite{hwang-noh}.

$f(R)$ Lagrangians in the gravitational action, conceived as
analytic functions of the Ricci scalar, deserve a special
attention. These actions are the simplest  among those which can
be constructed by  the curvature tensor invariants and their
covariant derivatives \cite{woodard}; they allow to determine
stable models with respect to small perturbations around maximally
symmetric solutions and to avoid ghost-like behaviors in effective
field theory. Beside cosmology,  such models could have
interesting consequences at  galactic and Solar System scales.
 In fact, although still controversial, the PPN limit of several $f(R)$
 theories of gravity could
be reconciled with the gravitational experiments, furnishing a
theoretical background for some measured deviation from GR
\cite{d,ppn-fR}. However, these results need a deeper
investigation and such a purpose will be pursued  together some
gravity dedicated experiments \cite{yunes}. However it is worth
stressing that $f(R)$ theories, in particular their simplest power
law form $f(R)=f_0R^n$, represent a viable model even from the
astrophysical point of view. It is in fact possible to demonstrate
that they can furnish a natural mechanism to interpret the
astrophysical data at galactic scales. Actually, thanks to the
corrections to the Newton potential which they give rise  in the
low energy limit \cite{noi-prl,noi-mnras}, one could achieve a
geometrical explanation to the missing matter problem (as
suggested also by other alternative approaches as MOND
\cite{mond}, which recently have been related to higher order
gravity itself \cite{navarro}.

An important issue regarding fourth order/modified gravity is
represented by its peculiarity to be recast in a
(non)\,-\,minimally coupled scalar-tensor theory by means of a
conformal transformation, namely, from the Jordan frame to the
Einstein one. In this way, in principle, one can pass from higher
order gravity to scalar-tensor gravity and vice-versa.
  From a  formal point of view,
this procedure is absolutely plausible while its physical
relevance is still controversial, \cite{cho,capozmarino,faraoni}
and, up to now, there is no privileged point of view regarding the
physical relevance of conformal transformations which remains
still debatable \cite{faraoni}.

Following standard approaches,  one cannot conclude univocally
about the physical relevance of the results. In general, it is
customary to pursue conformal transformations with the aim of
studying a physical problem developed in the Jordan frame and then
mapping it in the Einstein frame where the equations turn out to
be mathematically simpler. Therefore, Einstein frame equations are
solved and the solutions are considered as physically significant
\cite{crooks,kaloper,faraoniprd}. Despite  several results
achieved along this line, in order to get the physical meaning of
the problem, a back mapping of the results obtained in the
Einstein frame to the original Jordan one (see i.e.
\cite{hwang-conf,hwang-jord}) has to be in order. In general, the
back-transformation to the Jordan frame does not assure to get
well endowed physical solutions. It has been even widely
demonstrated that passing from one frame to the other can
completely change the physical meaning as well as the stability
 of the solutions one is dealing with \cite{sta}. For example a
scalar field quintessence model can roll a $\tilde{V}(\tilde\phi)$
potential motivated by particle physics in the Einstein frame,
while its back-transformation to the Jordan frame corresponds to a
potential $V(\phi)$ with no physical interest. Furthermore, in the
same sense, if a solution is accelerating in one frame, its
conformally transformed counterpart in the other frame is not
necessarily accelerating too \cite{faraoniprd}, or equivalently,
if one obtains a phantom model in the Einstein scheme, its mapping
in the Jordan frame does not give, in general, a phantom-like
evolution \cite{capoz-odintsov}. For example, if one takes into
account a conformally coupled scalar field with the potential
$\tilde{V}(\tilde\phi)\,=\,\lambda\tilde\phi^4$ in the Einstein
frame \cite{faraoniprd}, the conformal anti-transformation of such
a potential  gives\,:
\begin{equation}
V(\phi)\,=\,\left(\frac{3}{2k}\right)^2\lambda
\left(1-\frac{k}{6}\phi^2\right)^2\ln^4{\frac{\sqrt{k/6}\phi+1}{\sqrt{k/6}\phi-1}}\,,
\end{equation}
which is difficult to be physically interpreted. Such a problem is
particularly relevant if standard matter is considered in the
field equations due to non-minimal couplings arising thanks to the
conformal transformations. As an example, there is a recent result
where a  solution obtained from $f(R)$-gravity, transformed in the
Einstein frame and connecting the radiation epoch to the
accelerating attractor, $\tilde{a}\sim\tilde{t}^{3/5}$, is
considered non-physically relevant since  its back transformation
in the Jordan frame gives $a(t)\,=\,t^{1/2}$ which does not accord
with the expected dust matter dominated regime \cite{amendola}.
The interpretation of the result is  difficult and controversial
since assuming {\it a priori}  the Einstein frame as "physical",
we should rule out $f(R)$-theories. This means that, despite
several positive results achieved from extended modified theories
of gravity, without some hints coming from the original framework
in which the theory is conceived there is no way for a clear
physical interpretation of what is going on. In other words, is it
correct to obtain a solution in a frame and then  interpret it in
another frame?

In this paper,  we discuss  how extended gravity (in particular
$f(R)$-gravity) can be dealt under the standard of an ideal fluid
description or transformed to a (mathematically) equivalent
scalar-tensor theory. We show that the three descriptions could
not be physically equivalent. For example, the phantom phase in
one of the three approaches can corresponds to non-phantom phase
into another one. Furthermore, it is possible to show that a
dust-matter regime can be obtained in  a $f(R)$-gravity scenario,
and, in particular, its evolution can be connected to a late time
stage of accelerating expansion, without considering conformal
transformations. In Appendix it is demonstrated how any realistic
cosmology may be realized via the reconstruction of modified gravity.
These facts point out that the mathematical
equivalence among Jordan, Einstein and "fluid" descriptions does
not necessarily implies physical equivalence and solutions should
be carefully studied into the frames in which they are obtained.

\

\noindent 2. In this section, we  show that modified gravity can
be transformed also to GR with standard ideal fluid, so actually
two Einstein frames exist!

Let us start from the action of the $f(R)$-gravity: \be
\label{FR1} S=\int d^4 x \sqrt{-g}f(R)\ . \ee Here $f(R)$ is a
function of the scalar curvature $R$. For example, \be \label{FR2}
f(R)=\frac{1}{2\kappa^2}R - \frac{c_1}{R^n}\ , \ee is a
generalization of the model in \cite{CDTT}. In (\ref{FR2}), as
long as $n>-1$, $n$ is not restricted to be an integer but can be
an arbitrary real number. As a more generalized model, we may
consider\cite{NOPRD} \be \label{FR2b} f(R)=\frac{1}{2\kappa^2}R -
\frac{c_1}{R^n} + c_2 R^m\ , \ee Here we assume $m>1$. When the
curvature is small, the behavior of the model (\ref{FR2b}) is not
changed much from that of (\ref{FR2b}) but the model may describe
the inflation in the early universe. In the following, we assume
$c_{1,2}>0$.

For the model (\ref{FR1}), Starobinsky has proposed a condition
\be \label{Str1} \frac{d^2 f(R)}{dR^2}>0\ , \ee which could be
necessary for the stability of FRW solutions and for the existence
of the positive mass squared for the ``scalaron''. For the model
(\ref{FR2}), the condition gives \be \label{Str2}
-\frac{n(n+1)c_1}{R^{n+2}}>0\ , \ee which could be satisfied if
$-1<n<0$ when $c_1>0$ and $R>0$. And for the model (\ref{FR2b}),
we find \be \label{Str3} -\frac{n(n+1)c_1}{R^{n+2}} + m(m-1)c_2
R^{m-2}>0\ . \ee Then if $c_1$, $c_2$, $R>0$, we find \be
\label{Str4} c_2>\frac{n(n+1)c_1}{m(m-1)R^{n+m}}\ . \ee The
condition is satisfied again if $-1<n<0$. When $n>0$ (we do not
consider the case $n=0$ since it corresponds to the cosmological
constant), the condition (\ref{Str4}) is satisfied if $c_2$ is
large enough. Since the condition (\ref{Str4}) depends on the
curvature, even if the condition could be satisfied in the present
universe, when the curvature becomes smaller, the condition
(\ref{Str4}) may be violated. This may indicate that even more
terms should be considered in modified gravity action in the
future universe. Currently, such terms may be next-to-leading
order but they may become essential at extremely small curvature.

Technically, the action (\ref{FR1}) can be rewritten in the form of the scalar-tensor theory.
By introducing the auxiliary fields, $A$ and
$B$, one can rewrite the action (\ref{FR1}) as follows:
\be
\label{RR2b}
S=\int d^4 x \sqrt{-g} \left\{B\left(R-A\right) + f(A)\right\} \ .
\ee
One is able to eliminate $B$, and to obtain
\be
\label{RR6b}
S=\int d^4 x \sqrt{-g} \left\{f'(A)\left(R-A\right) + f(A)\right\}\ ,
\ee
and by using the conformal transformation $g_{\mu\nu}\to \e^\sigma g_{\mu\nu}$
$\left(\sigma = -\ln \left(2\kappa^2 f'(A)\right)\right)$, the action (\ref{RR6b}) is rewritten
as the Einstein-frame action:
\be
\label{RR10}
S_E=\frac{1}{2\kappa^2}\int d^4 x \sqrt{-g} \left\{ R - {3 \over 2}g^{\rho\sigma}
\partial_\rho \sigma \partial_\sigma \sigma - U(\sigma)\right\} \ .
\ee
Here,
\bea
\label{RR11b}
U(\sigma) &=& \e^\sigma G\left(\e^{-\sigma}\right)
 - 2\kappa^2 \e^{2\sigma} f\left(G\left(\e^{-\sigma} \right)\right) \nn
&=& {A \over 2\kappa^2 f'(A)} - {f(A) \over 2\kappa^2 f'(A)^2}\ ,
\eea
and $G$ is defined by $G\left(2\kappa^2 f'(A)\right)=A$, that is the inverse function
of $2\kappa^2 f'$.
In case of (\ref{FR2}), we find
\be
\label{FR4}
U(\sigma)=(n+1)\left(2\kappa^2 a\right)^{\frac{1}{n+1}}\e^{2\sigma}
\left(\e^{-\sigma} - 1\right)^{\frac{n}{n+1}}\ .
\ee

In principle, the inverse process is possible, that is, by
starting from the action of the Einstein gravity coupled with
scalar field, one can obtain the $f(R)$ action. By writing \be
\label{RR12b} \varphi = \pm
\frac{1}{\kappa}\sqrt{\frac{3}{2}}\sigma \ ,\quad \tilde
V(\varphi)=\frac{1}{2\kappa^2}U\left(\pm \kappa
\sqrt{\frac{2}{3}}\varphi\right)\ , \ee we can start from the
following action: \be \label{f6} S=\int d^4 x
\sqrt{-g}\left\{\frac{1}{2\kappa^2}R - \frac{1}{2}\partial_\mu
\varphi \partial^\mu \varphi  - \tilde V(\varphi)\right\}\ . \ee
Then one can use the back conformal transformation \be \label{f7}
g_{\mu\nu}\to \e^{\pm\kappa \varphi\sqrt{\frac{2}{3}}}g_{\mu\nu}\
, \ee and make the kinetic term of $\varphi$ vanish. Hence, one
obtains \be \label{f8} S=\int d^4 x
\sqrt{-g}\left\{\frac{\e^{\pm\kappa
\varphi\sqrt{\frac{2}{3}}}}{2\kappa^2}R
 - {\e^{\pm 2\kappa \varphi\sqrt{\frac{2}{3}}}}\tilde V(\varphi)\right\}\ .
\ee Since $\varphi$ becomes the auxiliary field, one can delete
$\varphi$ by using as an equation of motion: \be \label{f9}
R=\e^{\pm\kappa \varphi\sqrt{\frac{2}{3}}} \left(4\kappa^2 \tilde
V(\varphi) \pm 2\kappa \sqrt{\frac{3}{2}} \tilde
V'(\varphi)\right)\ , \ee which can be solved with respect to
$\varphi$ as $\varphi=\varphi(R)$. The action (\ref{f8}) is
re-incarnated in the form of $f(R)$-gravity, that is: \bea
\label{f10} S&=&\int d^4 x \sqrt{-g}f(R)\ , \nn f(R) &\equiv&
\frac{\e^{\pm \kappa \varphi(R)\sqrt{\frac{2}{3}}}}{2\kappa^2}R
\nn &&  - {\e^{\pm 2\kappa \varphi(R)\sqrt{\frac{2}{3}}}}\tilde
V\left(\varphi(R)\right)\ . \eea Therefore $f(R)$-gravity theory
is mathematically equivalent to the scalar-tensor theory which
does not yet mean their physical equivalence
\cite{capoz-odintsov}.

Let $W$ be the inverse function of  $\tilde V(\varphi)$: $W(\tilde
V(\varphi))=\varphi$. Then if we can solve the following
differential equation\footnote{ A trivial solution of (\ref{E1})
is $h(\phi)=h_0$ with a constant $h_0$. The solution corresponds
to de Sitter universe. We should note that $f(R)$-gravity, in
general, admits the de Sitter solution. The $h=h_0$ solution could
correspond to the de Sitter solution in $f(R)$-gravity. } \be
\label{E1} -\frac{1}{\kappa}\sqrt{-2\frac{dh(\phi)}{d\phi}} =
\frac{d}{d\phi}\left(W \left( \frac{1}{\kappa^2} \left( 3h(\phi)^2
+ \frac{dh(\phi)}{d\phi} \right) \right) \right)\ , \ee with
respect to $h(\phi)$, we can  further rewrite the action
(\ref{f6}) as \be \label{k1} S=\int d^4 x \sqrt{-g} \left\{
\frac{1}{2\kappa^2} R - \frac{1}{2} \omega(\phi)
\partial_\mu \phi\partial^\mu \phi  - V(\phi)\right\}\ .
\ee
Here
\be
\label{k6}
\omega(\phi)=- \frac{2}{\kappa^2}h'(\phi)\ ,\quad
V(\phi)=\frac{1}{\kappa^2}\left(3h(\phi)^2 + h'(\phi)\right)\ ,
\ee
and
\be
\label{f5}
\varphi=\int d\phi \sqrt{\left|\omega(\phi)\right|}\ .
\ee
For the action (\ref{k1}) with (\ref{k6}), if we can neglect
 contributions from other matter fields,
the following solution of the FRW equations  can be found
\cite{grg}: \be \label{k7} \phi=t\ ,\quad H=h(t)\ . \ee
Furthermore one finds the following, effective equation of state
(EoS): \be \label{SN1} p=-\rho -
\frac{2}{\kappa^2}h'\left(h^{-1}\left(\kappa\sqrt{\frac{\rho}{3}}\right)\right)\
, \ee Here $\rho$ and $p$ are the scalar energy density and the
pressure defined as: \be \label{k4} \rho =
\frac{1}{2}\omega(\phi){\dot \phi}^2 + V(\phi)\ ,\quad p =
\frac{1}{2}\omega(\phi){\dot \phi}^2 - V(\phi)\ . \ee We should
note that there is almost one to one (up to integration constants)
correspondence between $\tilde V(\varphi)$ in (\ref{k6}) and
$h(\phi)$, and therefore the corresponding EoS ideal fluid
(\ref{SN1}). Then for arbitrary $f(R)$-gravity, in principle, one
can find the corresponding ideal fluid EoS description (what maybe
considered as yet another Einstein frame).

As an example,  the small curvature case for (\ref{FR2}) may be
considered. Hence, \be \label{FFF1} U(\sigma) \sim
(n+1)\left(2\kappa^2 a\right)^{\frac{1}{n+1}}
\e^{\frac{n+2}{n+1}\sigma}\ , \ee or by using (\ref{RR12b}), \be
\label{FFF2} \tilde V(\varphi)= \frac{(n+1)\left(2\kappa^2
a\right)^{\frac{1}{n+1}}}{\kappa^2} \e^{\pm
\frac{n+2}{n+1}\kappa\sqrt{\frac{2}{3}}\varphi}\ . \ee In
(\ref{FFF2}), the two signs $\pm$ are independent with each other.
Then $W$ (\ref{E1}) has the following form: \be \label{FFF3}
W(x)=\pm \frac{n+1}{n+2}\frac{1}{\kappa}\sqrt{\frac{3}{2}} \ln
\left( \frac{\kappa^2}{(n+1)(2\kappa^2
a)^{\frac{1}{n+1}}}x\right)\ . \ee Here $x$ is an a-dimensional
variable, which is used to express the function $W$. Then a
non-trivial solution for (\ref{E1}) is given by \be \label{FFF4}
h(\phi)=3\left(\frac{n+1}{n+2}\right)^2\frac{1}{\phi}\ . \ee Using
(\ref{SN1}), we find an EoS with constant EoS parameter $w$: \be
\label{FFF5} p=w\rho\ ,\quad
w=\frac{2}{9}\left(\frac{n+2}{n+1}\right)^2 - 1\ . \ee For
instance, if \be \label{FFF6} n=-\frac{3 -
2\sqrt{3}}{3-\sqrt{2}}>-1\ , \ee we find $w=0$, which corresponds
to dust. On the other hand, if \be \label{FFF7} n=-\frac{\sqrt{3}
- 2}{\sqrt{3} - 1}>-1\ , \ee it follows $w=1/3$, which corresponds
to radiation. Hence, extended modified gravity may be presented in
mathematically equivalent form as GR with ideal fluid!

Note that as we have used the scale transformation, the cosmology appears
in a different way
in the frame corresponding to (\ref{FR2}) from the frame corresponding to (\ref{f6}). In the frame
corresponding to (\ref{f6}), which we call the Einstein frame, by using (\ref{k7}) and (\ref{FFF4}), we find
\be
\label{FFF8}
H=3\left(\frac{n+1}{n+2}\right)^2\frac{1}{t} \quad \mbox{or} \quad
a\propto t^{3\left(\frac{n+1}{n+2}\right)^2}\ ,
\ee
and by using (\ref{RR12b}), (\ref{f5}), and (\ref{f7}),
\be
\label{FFF9}
\sigma = \pm 2\left(\frac{n+1}{n+2}\right)^2\ln \frac{t}{t_0}\ .
\ee
Here $t_0$ is a constant of the integration in (\ref{f5}).
The time coordinate $t_J$ and the scale factor $a_J$ in the frame corresponding to (\ref{FR2}),
which we call as the Jordan frame, is related with $t$ and $a$ in the Einstein frame as
\bea
\label{FFF10}
&& dt_J=\e^{\sigma/2}dt \propto t^{\pm \left(\frac{n+1}{n+2}\right)^2}dt\ , \nn
&& a_J=\e^{\sigma/2} \propto t^{\pm \frac{n+1}{n+2} + 3\left(\frac{n+1}{n+2}\right)^2}\ .
\eea
Hence, one gets
\bea
\label{FFF11}
a_J&=&t_J^{h_J^\pm}\ ,\nn
h_J^\pm &=& \frac{(n+1)(4n+5)}{(2n+3)(n+2)}\ ,\quad
\frac{(n+1)(2n+1)}{n+2}\ .
\eea
The latter Eq.(\ref{FFF11}) reproduces the standard result.
We should note that the cosmology is really changed in the two frames.
 In the Einstein frame (\ref{FFF8}), we find
always $\dot H<0$, which corresponds to non-phantom phase.
On the other hand, in the Jordan frame,
 we find
\be \label{FFF12} H_J\equiv \frac{\dot a_J}{a_J} =
\frac{(n+1)(2n+1)}{n+2}\frac{1}{t}\ . \ee Then if $-1<n<-1/2$, it
follows $\dot H>0$, which corresponds to phantom phase. We should
note that in the phantom regime $-1<n<-1/2$, Starobinsky's
condition (\ref{Str1}) is satisfied. Hence, we demonstrated that
mathematical equivalence does not always means the physical
equivalence (the same point of view was recently expressed in
\cite{thomas}). In conclusion, we can state that the existence of
so many different frames indicates that  only one, the original
frame, can be considered the physical one.

\

\noindent 3. As we have seen, by performing a conformal
transformation on a modified gravity Lagrangian $f(R)$, it is
possible to achieve, in the Einstein frame, dust matter behaviours
which are compatible with observational prescriptions. In
addition, by exploiting the analogy between the two frames and
between modified gravity and scalar-tensor gravity one can realize
that physical results, in the two conformally related frames,
could be completely different. In other words one can pass from a
non\,-\,phantom phase behaviour (Einstein frame) to a phantom
regime (Jordan frame). Now, we can suppose to change completely
the point of view. In fact, we can rely directly with the Jordan
frame and we can verify if a dust matter regime is intrinsically
compatible with modified gravity.

As a first  example, one can cite the exact solution provided in
\cite{noi-ijmpd}, which has been deduced working solely in the
Jordan Frame (FRW universe). In particular, one is able to find a
power law regime for the scale factor whose rate is connected with
the power $n$ of the
 Lagrangian $f(R)\,=\,f_0 R^n$. In other words, one has
$a(t)\,=\,a_0 t^{\alpha}$ with
$\alpha\,=\,\frac{-2n^2+3n-1}{n-2}$. Such an exact solution is
found out when only baryonic matter is considered
\cite{noi-prl,noi-mnras}. It is evident that such a solution
allows to obtain an ordinary matter behaviour ($\alpha\,=\,2/3$)
for given values of the parameter $n$ (i.e. $n \sim -0.13,\ n \sim
1.29$). Such solutions  are nevertheless stable and no transition
to acceleration phase then occurs. In general, it is possible to
show that solutions of the type
\begin{equation}
\label{friedman-transient} a\,=\,a_0(t-t_0)^{\frac{2n}{3(1+w)}}\,,
\end{equation}
where  $w$ is the barotropic index of standard perfect fluid,
 arises as a transient phase, and this phase evolves into an
accelerated solution representing an attractor for the system
\cite{noi-phase}. In any case,  a single solution exactly
matching, in sequence, radiation, matter and accelerated phases is
unrealistic to be found out in  the framework of  simple
$f(R)$-power law theories. The discussion can be further extended.
Modified gravity can span a wide range of analytic functions of
the Ricci scalar where $f(R)\,=\,f_0R^n$ only represents  the
simplest choice. In general, one can reverse the perspective and
try to derive the form of  gravity Lagrangian directly from the
data or mimicking other cosmological models. Such an approach has
been developed in \cite{mimick}, and allows to recover modified
gravity Lagrangians by the Hubble flow dynamics $H(z)$: in
particular, it is possible to show that wide classes of  Dark
Energy models worked out in the Einstein frame can be consistently
reproduced by $f(R)$-gravity as
 quintessence models with exponential potential
\cite{mimick-scafi}. Clearly the approach works also for  the case
of coupled quintessence scalar field. In other words, the dynamics
of $H(z)$, considered in the Jordan frame, is reconstructed by
observational data considered in the Einstein frame then assuming
one of the two frames as the "physical frame" could be misleading.
Here we further develop this approach with the aim to show the
viability of $f(R)$ gravity to recover a matter-dominated phase
capable of evolving in a late accelerating phase.

 From a formal point of view, the reconstruction of the gravity
Lagrangian from data is based on the relation which expresses the
Ricci scalar in terms of the Hubble parameter\,:
\begin{equation}
R = -6 \left ( \dot{H} + 2 H^2 +\frac{k}{a^2}\right )\,. \label{eq: constr}
\end{equation}
Now, the modified gravity field equation reads\,
\cite{capoz-curv,noi-ijmpd}:
\begin{equation}
G_{\alpha \beta} = R_{\alpha \beta} -  \frac{1}{2} R g_{\alpha \beta} = T^{(curv)}_{\alpha \beta} +
T^{(m)}_{\alpha \beta} \label{eq: field}
\end{equation}
where a stress\,-\,energy  tensor summing up
 all the higher order contributions is defined. It plays the
role of an effective curvature fluid\,:
\begin{eqnarray}
T^{(curv)}_{\alpha \beta} & = & \frac{1}{f'(R)} \left \{ g_{\alpha \beta} \left [ f(R) - R f'(R) \right ] /2 +\right .
\nonumber \\
 ~ & ~ & \nonumber \\
 ~ & + & \left . f'(R)^{; \mu \nu} \left( g_{\alpha \mu} g_{\beta \nu} - g_{\alpha \beta} g_{\mu \nu} \right )
\right \} \label{eq: curvstress}
\end{eqnarray}
while the matter term enters Eq.(\ref{eq: field}) through  the
modified stress\,-\,energy tensor $T^{(m)}_{\alpha \beta} =
\tilde{T}^{(m)}_{\alpha \beta}/f'(R)$ with
$\tilde{T}^{(m)}_{\alpha \beta}$ the standard minimally coupled
matter stress\,-\,energy tensor. Starting from this general
scheme, one can reconstruct the form of $f(R)$ from the Hubble
parameter as a function of the redshift $z$ exploiting the
relation (\ref{eq: constr}) after this expression has been
rewritten in term of the redshift itself. A key role in this
discussion is played by the conservation equation for the
curvature and the matter fluids which, in the case of dust matter,
(i.e. $p_m = 0$) gives\,:
\begin{eqnarray}
\dot{\rho}_{curv} + 3 H (1 + w_{curv}) \rho_{curv} & = & - \frac{1}{f'(R)} (\dot{\rho}_m + 3 H \rho_m) \nonumber \\
 ~ & ~ & - \rho_m \frac{df'(R)}{dt} \ . \label{eq: cons}
\end{eqnarray}
In particular, one may assume that the matter energy density is
conserved\,:
\begin{equation}
\rho_m = \Omega_M \rho_{crit} a^{-3} = 3 H_0^2 \Omega_M (1 + z)^3 \label{eq: mattrho}
\end{equation}
with $z = 1/a - 1$ the redshift   (having set $a(t_0) = 1$),
$\Omega_M$ the matter density parameter (here and hereon
quantities labelled with the subscript $0$ refers to present day
($z = 0$) values). Eq.(\ref{eq: mattrho}) inserted into
Eq.(\ref{eq: cons}), allows to write a conservation equation for
the effective curvature fluid\,:
\begin{eqnarray}
\dot{\rho}_{curv} + 3 H (1 + w_{curv}) \rho_{curv} & = & 3 H_0^2 \Omega_M (1 + z)^3  \nonumber \\
 ~ & ~ & \times \ \frac{\dot{R} f''(R)}{\left [ f'(R) \right ]^2} \ . \label{eq: curvcons}
\end{eqnarray}
Actually, since the continuity   equation and the field equations
are not independent  \cite{mimick}, one can reduce  to the
following single equation
\begin{eqnarray}
\dot{H} & = & -\frac{1}{2 f'(R)} \left \{ 3 H_0^2 \Omega_M (1 + z)^3 + \ddot{R} f''(R)+ \right . \nonumber \\
 ~ & ~ & \left . + \dot{R} \left [ \dot{R} f'''(R) - H f''(R) \right ] \right \} \,, \label{eq: presingleeq}
\end{eqnarray}
where all quantities can be expressed in term of  redshift by
means of the relation $\frac{d}{dt} = - (1 + z) H \frac{d}{dz}$.
In particular, for a flat FRW metric, one has\,:
\begin{equation}
R = -6 \left [ 2 H^2 - (1 + z) H \frac{dH}{dz} \right ] \,,\label{eq: rvsh}
\end{equation}
\begin{equation}
f'(R) = \left ( \frac{dR}{dz} \right )^{-1} \frac{df}{dz} \ , \label{eq: fp}
\end{equation}
\begin{equation}
f''(R) = \left ( \frac{dR}{dz} \right )^{-2} \frac{d^2f}{dz^2} - \left ( \frac{dR}{dz} \right )^{-3}
\frac{d^2R}{dz^2} \frac{df}{dz} \ , \label{eq: fpp}
\end{equation}
\begin{eqnarray}
f'''(R) & = & \left ( \frac{dR}{dz} \right )^{-3} \frac{d^3f}{dz^3}
+ 3 \left ( \frac{dR}{dz} \right )^{-5} \left ( \frac{d^2R}{dz^2} \right )^2 \frac{df}{dz} +\nonumber \\
 ~ & - & \left ( \frac{dR}{dz} \right )^{-4} \left ( 3 \frac{d^2R}{dz^2} \frac{d^2f}{dz^2} + \frac{d^3R}{dz^3}
\frac{df}{dz} \right ) \ . \label{eq: f3p}
\end{eqnarray}
Now, we have all the ingredients to reconstruct the shape of
$f(R)$ by data or, in general,  by the definition of a suitable
$H(z)$ viable with respect to observational results. In
particular,  we can show that a standard matter regime (necessary
to cluster large scale structure) can arise, in this scheme,
before the accelerating phase arises as in the so called  {\it
quiessence} model. A quiessence model is based on an ordinary
matter fluid plus a cosmological component whose equation of state
$w$ is constant but can scatter from  $w\,=\,-1$. This approach
represents the easiest generalization of the cosmological constant
model, and it has been successfully tested against the SNeIa
Hubble diagram and the CMBR anisotropy spectrum so that it allows
to severely constraint the barotropic index $w$ \cite{EstW}. It is
worth noticing that these constraints extend into the region $w <
-1$ therefore models (phantom models) violating the weak energy
condition are allowed. From the cosmological dynamics viewpoint,
such a model, by definition, has to display an evolutionary rate
of expansion which moves from the standard matter regime to the
accelerated behaviour in relation to the value of $w$. In
particular, this quantity parameterizes  the transition point to
the accelerated epoch. Actually, if it is possible to find out a
$f(R)$-gravity model compatible with the evolution of the Hubble
parameter of the quiessence model, this result suggests that
modified gravity  is compatible with a phase of standard matter
domination. To be precise, let us consider the Hubble flow defined
by this model\,:
\begin{equation}
H(z) = H_0 \sqrt{\Omega_M (1 + z)^3 + \Omega_X (1 + z)^{3 (1 + w)}} \label{eq: hqcdm}
\end{equation}
with $\Omega_X = (1 - \Omega_M)$ and $w$ the constant parameter
defining the dark energy barotropic index. This definition of the
Hubble parameter implies\footnote{Note that $R$ is always negative
as a consequence of the signature $\{+, -, -, -\}$ adopted. If we
had used the opposite signature, Eq.(\ref{eq: rvszqcdm}) is the
same, but with an overall positive sign.}\,:
\begin{equation}
R = - 3 H_0^2 \left [ \Omega_M (1 + z)^3 + \Omega_X (1 - 3 w) (1 + z)^{3 (1 + w)} \right ] \,. \label{eq:
rvszqcdm}
\end{equation}
The ansatz in Eq.(\ref{eq: hqcdm}) allows to obtain from
Eq.(\ref{eq: presingleeq}) a differential relation for $f(R(z))$
which can be solved numerically by choosing suitable boundary
 conditions. In particular we choose\,:
\begin{equation}
\left ( \frac{df}{dz} \right )_{z = 0} = \left ( \frac{dR}{dz} \right )_{z = 0} \ , \label{eq: fpzero}
\end{equation}
\begin{equation}
\left ( \frac{d^2f}{dz^2} \right )_{z = 0} = \left ( \frac{d^2R}{dz^2} \right )_{z = 0} \ . \label{eq: fppzero}
\end{equation}
\begin{equation}
f(z = 0) = f(R_0) = 6 H_0^2 (1 - \Omega_M) + R_0 \,. \label{eq: fzero}
\end{equation}
A comment  is in order here.  We have derived the present day
values of $df/dz$ and $d^2f/dz^2$ by imposing the consistency of
the reconstructed $f(R)$ theory with {\it local} Solar System
tests. One could wonder whether tests on local scales could be
used to set the boundary conditions for a cosmological problem. It
is easy to see that this is indeed meaningful. Actually, the
isotropy and homogeneity of the universe ensure that the present
day value of a whatever cosmological quantity does not depend on
where the observer is. As a consequence,  hypothetical observers
living in the Andromeda galaxy and testing gravity in his
planetary system should get the same results. As such, the present
day values of $df/dz$ and $d^2f/dz^2$ adopted by these
hypothetical observers are the same as those we have used based on
our Solar System experiments. Therefore, there is no systematic
error induced by our method of setting the boundary conditions.

Once one has obtained  the numerical solution for $f(z)$,
inverting again numerically Eq.(\ref{eq: rvszqcdm}), we may obtain
$z = z(R)$ and finally get $f(R)$  for several values of $w$. It
turns out that $f(R)$ is the same for different models for low
values of $R$ and hence of $z$. This is a consequence of the well
known degeneracy among different quiessence models at low $z$
that, in the standard analysis, leads to large uncertainties on
$w$.  This is reflected in the shape of the reconstructed $f(R)$
that is almost $w$\,-\,independent in this redshift range. An
analytic representation of the reconstructed fourth order gravity
model, can be obtained considering that the following empirical
function

\begin{equation}
\ln{(-f)} = l_1 \left [ \ln{(-R)} \right ]^{l_2} \left [ 1 + \ln{(-R)} \right ]^{l_3} + l_4 \label{eq: fit}
\end{equation}
approximates very well the numerical  solution, provided that the
parameters $(l_1, l_2, l_3, l_4)$ are suitably chosen for a given
value of $w$. For instance, for $w = -1$ (the cosmological
constant) it is\,:
\begin{displaymath}
(l_1, l_2, l_3, l_4) = (2.6693, 0.5950, 0.0719, -3.0099) \ .
\end{displaymath}
At this point, one can wonder if it is possible to improve such a
result considering even the radiation, although energetically
negligible. Rather than inserting radiation in the (\ref{eq:
hqcdm}), a more general approach in this sense is to consider the
Hubble parameter descending from a unified model like those
discussed  in \cite{unified}. In such a scheme one takes into
account energy density which scales as\,:
\begin{equation}
\rho(z) = A \ \left ( 1 + \frac{1 + z}{1 + z_s} \right )^{\beta - \alpha} \ \left [ 1 + \left ( \frac{1 + z}{1 +
z_b} \right )^{\alpha} \right ] \label{eq: rhoz}
\end{equation}
having defined\,:
\begin{equation}\label{eq: defzb}
z_s = 1/s - 1 \ ,\ \ \ \ \ \ \ \ z_b = 1/b - 1 \ .
\end{equation}
This model, with the choice $(\alpha, \beta) = (3, 4)$,  is able
to mimic a universe undergoing first a radiation dominated era
(for $z \gg z_s$), then a matter dominated phase (for $z_b \ll z
\ll z_s$) and finally approaching a de\,Sitter phase with constant
energy. In other words it works  in the way we are asking for. In
such a case, the Hubble parameter can be written, in natural
units, as $H\,=\,\sqrt{\frac{\rho(z)}{3}}$ and one can perform the
same calculation as in the quiessence case.
 As final result, it is  again possible to find out a suitable
$f(R)$-gravity model which,  for numerical reasons, it is
preferable to interpolate as $f(R)/R$\,:
\begin{displaymath}\label{fr-hobbit}
\frac{f(R)}{R}\,=\,1.02\times\frac{R}{R_0}\left[1+\left(-0.04\times(\frac{R}{R_0})^{0.31} \right.\right.
\end{displaymath}
\begin{equation}
\left.\left. +0.69\times(\frac{R}{R_0})^{-0.53}\right)\times \ln({\frac{R}{R_0}})\right] \,,
\end{equation}
where $R_0$ is a normalization constant. This result once more
confutes issues addressing modified gravity as incompatible with
structure formation prescriptions. In fact, also in this case, it
is straightforward to show that a phase of ordinary matter
(radiation and dust) domination  can be obtained and it is
followed by an accelerated phase.

\

\noindent 4. In this paper we have discussed the possibility, in
the framework of $f(R)$-gravity models, to obtain a standard
matter phase  followed by accelerated expansion.  We have faced
this task following two different lines: in a first case, assuming
a gravity Lagrangian $ \propto R-c_1 R^{-n}$ we have outlined the
mapping of $f(R)$ gravity from the Jordan Frame to the Einstein
one by means of a conformal transformation. We have showed that,
as matter of fact, dust matter domination is possible in this
approach and a late time accelerating phase naturally arises. When
a minimally coupled scalar field Lagrangian is mapped back into
the Jordan frame, one can deduce that modified gravity can imply
an effective fluid contribution whose origin is the geometrical
counterpart. The consequent EoS (\ref{FFF5}) shows a constant
barotropic index which, in relation to the choice of the model
parameter $n$, can provide behaviours compatible with ordinary
matter evolutions. Furthermore, such a parameterization allows
also to describe  phantom-like regimes, which turn out to be
inspired by $f(R)$-gravity. When a mapping into the Einstein frame
is considered, this solution does not give $\dot H
> 0$ so that the phantom\,-\,like behaviour is lost.
This means that the conformal transformation does not necessarily
imply physical equivalence of solutions and this difference among
the two frames  could represent a pathology of back mapping
conformal frame solutions. The debate on this point is recently
growing and several studies have been pursued in this direction
(see e.g.  \cite{amendola}). In order to avoid pathologies related
to the conformal transformation of solutions, another approach can
be pursued, discussing the viability of dust matter regimes
directly in the Jordan Frame. We have exploited a reconstruction
procedure of $f(R)$-gravity by means of the phenomenological (or
observational) knowledge of the Hubble parameter in term of the
redshift. The $H(z)$ rate can be inspired by data fitting of
speculative models like Q-essence or by means of unified schemes
considering an ordinary matter-dominated expansion followed by a
late time acceleration. Since in both cases $f(R)$-gravity  is
compatible with the assumed $H(z)$, one can conclude that modified
gravity is sufficiently versatile to furnish the dynamics of a
viable cosmic evolution.

In summary, it seems that $f(R)$-models are viable and cannot be
ruled out only assuming mathematical equivalence of conformal
transformations. However, considering simple $f(R)=f_0 R^n$ models
cannot completely solve the cosmic evolution problem since
matter-dominated solutions  for $n\neq 1$ are generally stable and
do not give rise to late-time transitions to accelerated behaviors
(see also \cite{amendola1}). The situation is similar to that
faced several times in inflationary cosmology  where graceful exit
from inflation regime can result problematic for simple models. In
other words, $f(R)$ gravity could be the way to bypass
shortcomings as dark energy and dark matter at cosmological and
astrophysical scales taking into account only {\it observed}
ingredients  as baryons, radiation and gravity but more realistic
Lagrangians have to be taken into account than simple power-law
ones.

 In the following appendix, we
show that it is always possible to work out a $f(R)$-model capable
of matching with cosmological radiation, matter and accelerating
epochs in succession.

\tolerance=5000 {\bf Appendix}. Let us consider the reconstruction
of modified gravity. That is, first we consider the proper Hubble
rate $H$, which describes the evolution of the universe, with
radiation dominance, matter dominance, and accelerating expansion.
It turns out that one can find $f(R)$-theory realizing such a
cosmology (with or without matter). The construction is not
explicit and we need to solve the second order differential
equation and algebraic equation. It shows, however, that, at
least, in principle, we could obtain any cosmology by properly
reconstructing a function $f(R)$ on theoretical level.

The starting action is \be \label{PQR1} S=\int d^4 x
\left\{P(\phi) R + Q(\phi) + {\cal L}_{\rm matter}\right\}\ . \ee
Here $P$ and $Q$ are proper functions of the scalar field $\phi$
and ${\cal L}_{\rm matter}$ is the matter Lagrangian density.
Since the scalar field does not have a kinetic term, we may regard
$\phi$ as an axilliary field. In fact, by the variation of $\phi$,
it follows \be \label{PQR2} 0=P'(\phi)R + Q'(\phi)\ , \ee which
may be solved with respect to $\phi$: \be \label{PQR3}
\phi=\phi(R)\ . \ee By substituting (\ref{PQR3}) into
(\ref{PQR1}), one obtains $f(R)$-gravity: \bea \label{PQR4}
S&=&\int d^4 x \left\{f(R) + {\cal L}_{\rm matter}\right\}\ , \nn
f(R)&\equiv& P(\phi(R)) R + Q(\phi(R))\ . \eea

By the variation of the action (\ref{PQR1}) with respect to the
metric $g_{\mu\nu}$, we obtain \bea \label{PQR5}
0&=&-\frac{1}{2}g_{\mu\nu}\left\{P(\phi) R + Q(\phi) \right\} -
R_{\mu\nu} P(\phi) \nn && + \nabla_\mu \nabla_\nu P(\phi) -
g_{\mu\nu} \nabla^2 P(\phi) + \frac{1}{2}T_{\mu\nu}\ . \eea FRW
equations are \bea \label{PQR6}
0&=&-6 H^2 P(\phi) - Q(\phi) - 6H\frac{dP(\phi(t))}{dt} + \rho \ ,\\
\label{PQR7} 0&=&\left(4\dot H + 6H^2\right)P(\phi) + Q(\phi) \nn
&& + 2\frac{d^2 P(\phi(t))}{dt} + 4H\frac{d P(\phi(t))}{dt} + p\ .
\eea By combining (\ref{PQR5}) and (\ref{PQR6}) and deleting
$Q(\phi)$, we find the following equation \be \label{PQR7b} 0=2
\frac{d^2 P(\phi(t))}{dt^2} - 2 H \frac{dP(\phi(t))}{d\phi} +
4\dot H P(\phi) + p + \rho\ . \ee As one can redefine the scalar
field $\phi$ properly, we may choose \be \label{PQR8} \phi=t\ .
\ee We also assume, as matter, for example, a combination of the
radiation and dust: \be \label{PQR9} \rho=\rho_{r0} a^{-4} +
\rho_{d0} a^{-3}\ ,\quad p=\frac{\rho_{r0}}{3}a^{-4}\ . \ee Here
$\rho_{r0}$ and $\rho_{d0}$ are constants. If the scale factor $a$
is given by a proper function $g(t)$ as \be \label{PQR10}
a=a_0\e^{g(t)}\ , \ee with a constant $a_0$, Eq.(\ref{PQR7})
reduces to the second rank differential equation: \bea
\label{PQR11} 0&=&2 \frac{d^2 P(\phi)}{d\phi^2} + 4g''(\phi)
P(\phi) - 2 g'(\phi) \frac{dP(\phi))}{dt} \nn && +
\frac{4}{3}\rho_{r0} a_0^{-4}\e^{-4g(\phi)} + \rho_{d0}
a_0^{-3}\e^{-3g(\phi)}\ . \eea In principle, by solving
(\ref{PQR11}) we find the form of $P(\phi)$. Then by using
(\ref{PQR6}) (or equivalently (\ref{PQR7})), we also find the form
of $Q(\phi)$ as \bea \label{PQR12} Q(\phi)&=&-6
\left(g'(\phi)\right)^2 P(\phi) - 6g'(\phi) \frac{dP(\phi)}{d\phi}
\nn && + \rho_{r0} a_0^{-4}\e^{-4g(\phi)} + \rho_{d0}
a_0^{-3}\e^{-3g(\phi)}\ . \eea Then in principle, any cosmology
expressed as (\ref{PQR10}) can be realized by $f(R)$-gravity.

For example, we consider the case \be \label{PQR13} g'(\phi)=g_0 +
\frac{g_1}{\phi}\ , \ee and without matter $\rho=p=0$ for
simplicity. Then Eq.(\ref{PQR11}) reduces as \be \label{PQR14}
0=\frac{d^2 P}{d\phi^2} - \left(g_0 +
\frac{g_1}{\phi}\right)\frac{dP}{d\phi}
   - \frac{2g_1}{\phi^2}P\ ,
\ee whose solutions are given by the Kummer function
(hypergeometric function of confluent type) as \be \label{PQR15}
P=z^\alpha F_K(\alpha,\gamma; z)\ , \quad z^{1-\gamma} F_K(\alpha
- \gamma + 1, 2 - \gamma; z)\ . \ee Here \bea \label{PQR16} &&
z\equiv g_0\phi \ ,\quad \alpha\equiv \frac{1+g_1 \pm \sqrt{g_1^2
+ 2g_1 + 9}}{4}\ ,\nn && \gamma\equiv 1\pm \frac{\sqrt{g_1^2 +
2g_1 + 9}}{2}\ , \eea and the Kummer function is defined by \be
\label{PQR17} F_K(\alpha,\gamma;z)=\sum_{n=0}^\infty
\frac{\alpha(\alpha + 1)\cdots (\alpha + n -1)}{\gamma(\gamma +
1)\cdots (\gamma + n - 1)} \frac{z^n}{n!}\ . \ee Eq.(\ref{PQR13})
tells that the Hubble rate $H$ is given by \be \label{PQR18} H=g_0
+ \frac{g_1}{t}\ . \ee Then when $t$ is small, as $H\sim g_1/t$,
the universe is filled with a perfect fluid with the EoS parameter
$w=-1 + 2/3g_1$. On the other hand when $t$ is large $H$
approaches to constant $H\to g_0$ and the universe looks as
deSitter space. This shows the possibility of the transition from
matter dominated phase to the accelerating phase. Similarly, one
can construct modified gravity action describing other epochs
bearing in mind that form of the modified gravity action is
different at different epochs (for instance, in inflationary epoch
it is different from the form at late-time universe).

We now investigate the asymptotic forms of $f(R)$ in (\ref{PQR4})
corresponding to (\ref{PQR13}). When $\phi$ and therefore $t$ are
small, we find \be \label{PQR19} P\sim P_0\phi^\alpha\ ,\quad
Q\sim -6P_0g_1\left(g_1 + \alpha\right)\phi^{\alpha - 2}\ . \ee
Here $P_0$ is a constant. Then by using (\ref{PQR2}), we find \be
\label{PQR20} \phi^2 \sim \frac{6g_1\left(g_1 +
\alpha\right)\left(\alpha - 2\right)}{\alpha R}\ , \ee which gives
\be \label{PQR21} f(R) \sim - \frac{2P_0}{\alpha - 2}\left\{
\frac{6g_1\left(g_1 + \alpha\right)\left(\alpha -
2\right)}{\alpha}\right\}^{\alpha/2} R^{1 - \frac{\alpha}{2}}\ .
\ee On the other hand, when $\phi$ and therefore $t$ are positive
and large, one gets \bea \label{PQR22} P&\sim & \tilde P_0
\phi^{2\alpha - \gamma}\e^{g_0\phi}\left(1 +
\frac{(1-\alpha)(\gamma - \alpha)}{g_0\phi}
%+ {\cal O}\left(\phi^{-2}\right)
\right) \nn Q&\sim & - 12 g_0^2 \tilde P_0 \phi^{2\alpha -
\gamma}\e^{g_0\phi} \nn && \times \left(1 + \frac{-9 + 12 \alpha -
5\gamma - 2\alpha \gamma + 2\alpha^2}{2g_0\phi}
%+ {\cal O}\left(\phi^{-2}\right)
\right) \nn \phi&\sim& \frac{-\frac{9}{2} + 9\alpha -
\frac{7}{2}\gamma}{g_0\left(\frac{R}{12g_0^2} - 1\right)}\ . \eea
Here $\tilde P_0$ is a constant. Then we find \bea \label{PQR23}
f(R)&\sim& 12 g_0^2 \tilde P_0 \left\{\frac{1}{g_0} \left(
-\frac{9}{2} + 9\alpha - \frac{7}{2}\gamma
\right)\right\}^{2\alpha - \gamma} \nn && \times
\left(\frac{R}{12g_0^2} - 1\right)^{-2\alpha + \gamma + 1} \nn &&
\times \exp\left( \frac{ -\frac{9}{2} + 9\alpha -
\frac{7}{2}\gamma }{\frac{R}{12g_0^2} - 1}\right)\ . \eea This
shows the principal possibility of unification of matter-dominated
phase, transition to acceleration and late-time speed up.

\

\noindent {\bf Acknoweledgements.} We are grateful to A.
Starobinsky and V.F. Cardone for useful remarks which allowed to
improve the paper. We thank also L. Amendola and S. Tsujikawa for
discussions and comments on the topic.  The research by SDO was
supported in part by LRSS project n4489.2006.02(Russia), by RFBR
grant 06-01-00609 (Russia), by project FIS2005-01181(MEC, Spain).

\end{document}